\documentclass[hyper]{JHEP} 

    \usepackage{epsfig}

    \newcommand\fverb{\setbox\pippobox=\hbox\bgroup\verb}

    \newcommand\fverbdo{\egroup\medskip\noindent%

                            \fbox{\unhbox\pippobox}\ }

    \newcommand\fverbit{\egroup\item[\fbox{\unhbox\pippobox}]}

    \newbox\pippobox

    \title{New Solution of
     Open Bosonic 
    String Field Theory}

    \author{ J. Kluso\v{n}
    \footnote{On leave from Masaryk University, Brno}\\
    Institute of Theoretical Physics, University of Stockholm, SCFAB\\
    SE- 106 91 Stockholm, Sweden \\
    and \\
    Institutionen f\"or teoretisk fysik\\
    BOX 803, SE- 751 08 
    Uppsala, Sweden \\
    E-mail: \email{josef.kluson@teorfys.uu.se}}

    \preprint{\hepth{0205294}}

    \abstract{We present 
    example of exact   solution
    to Witten's open bosonic string field theory.
    We will analyse the new BRST operator and we
    will argue that the new solution describes the
    flow from zero slope  limit 
    ($\alpha'\rightarrow 0$)
    to the tensionless limit ($\alpha'\rightarrow
    \infty$) in the string world-sheet action.}
    \keywords{String field theory}

\begin{document}
    \section{Introduction}\label{first}
    Recently there has been great interest 
    open bosonic  string field theory
    \cite{Witten:1985cc}, especially
    in the context of the tachyon condensation
    (For a review of  string field theory, see
    \cite{Ohmori:2001am,DeSmet:2001af,Arefeva:2001ps,
    Siegel:1988yz}.) The remarkable sucess of string
    field theory in addressing this
    question leads us to believe that string
    field theory has a profound
    role to play in the formulation of string theory.
     
    In some previous papers we have proposed a method
    for finding exact solutions to the open bosonic
    string field theory  \cite{Kluson:2002kk,
    Kluson:2002ex,Kluson:2002hr}. This method relies
    on the existence of a ghost number zero operator
    that does not commute with the BRST operator so that
    the string field theory formulated around the
    new background corresponds to a changed BRST operator.

    In this short note we continue this investigation and 
    find new exact solutions to the open
    bosonic string field theory action and the corresponding
    modified BRST operator which have very
    unusual properties. We show that
    the form of the new BRST operator depends on a
    single parameter $t$  that can vary from $0$ to $\infty$.
    For $t=0$ the BRST operator 
    is the same as the BRST operator for a massless
    particle. For $t=1$ we have the original D25-brane,
    and finally, for $t=\infty $ the BRST operator
    corresponds to the BRST operator for a tensionless string.
    
    A crucial point is that the effective tension that
    we introduce during the calculation appears to effectively
    depend on the form of the string field theory
    solution depends on the direction where we allow
    our operator $K$ applies.

    This has the remarkable consequence
    in the emergence of a string with 
    an-isotropic string tension. A similar
    situation  has previously been  studied in
    the context of the noncommutative open string
    (NCOS) \cite{Seiberg:2000ms,Gopakumar:2000na,
    Gomis:2000zz,Aharony:2000gz,Gopakumar:2000ep,
    Bergshoeff:2000ai,Gomis:2000bd,
    Danielsson:2000gi,Klebanov:2000pp,
    Lu:2001qf,Danielsson:2000mu,Kristiansson:2000xv,
    Garcia:2002fa}. Our result
    deserves further study in that context.We believe our results merit further study. 
    Especially it would be nice to perform an
    explicit CFT analysis of a world-sheet theory
    with an-isotropic tension.

    \section{Review of the general
    formalism}\label{second}
    In this section we briefly review
    the general formalism introduced  in
    \cite{Kluson:2002kk} and then  applied in
    \cite{Kluson:2002ex,Kluson:2002hr}.

    The action of the open bosonic string 
    field theory \cite{Witten:1985cc}  
    has the form
    \begin{equation}\label{SFTaction}
    S=-\left(\frac{1}{2}\int \Phi\star Q \Phi+
    \frac{1}{3}\int \Phi\star \Phi\star
    \Phi \ \right).
    \end{equation}
    Varying the action leads to the equation of motion
    \begin{equation}\label{eqm}
    Q\Phi+\Phi\star \Phi=0 \ .
    \end{equation}
    As demonstrated in  \cite{Kluson:2002kk}, it  is easy  to see that 
    the field
  
    \begin{equation}\label{sol}
    \Phi_0=e^{-K_L(\mathcal{I})}\star Q
    (e^{K_L(\mathcal{I})})
    \end{equation}
     is a  solution of the equation of motion
    (\ref{eqm})
    since we have
    \begin{equation}
    Q  (\Phi_0)=Q(e^{-K_L(\mathcal{I})}
    \star Q(e^{K_L(\mathcal{I})}))=
    -e^{-K_L(\mathcal{I})}
    \star Q(e^{K_L(\mathcal{I})})\star 
    e^{-K_L(\mathcal{I})}
    \star   Q(e^{K_L(\mathcal{I})})
    \end{equation}
    and
    \begin{equation}
    \Phi_0\star \Phi_0=e^{-K_L(\mathcal{I})}
    \star Q(e^{K_L(\mathcal{I})})
    \star e^{-K_L(\mathcal{I})}
    \star Q(e^{K_L(\mathcal{I})}) \ .
    \end{equation}
    where $K$ is a ghost number zero operator  
    that obeys
    \begin{eqnarray}\label{Krule}
    K(X\star  Y)=K(X)\star Y+
    X\star K(Y) , \ \nonumber \\
    \int K(X)\star Y=-\int X  
    \star K(Y) \ ,
    \nonumber \\
    \end{eqnarray}
    and where $\mathcal{I}$ is the identity ghost number
    zero field \cite{Horowitz:dt} that obeys
    \begin{equation}
    \mathcal{I}\star \Psi=\Psi \star \mathcal{I}=
    \Psi \ .
    \end{equation}
    for any string field $\Psi$
    \footnote{A recent discussion of this special
    string field is given in \cite{Ellwood:2001ig,
    Kishimoto:2001de}}.
     In (\ref{sol}) the
    index $L$ means the integration of the local
    density over the left side of the string that is
    labeled by $\sigma\in (0,\pi/2)$. Similarly,
    the notation $K_R$ corresponds to the integration
    of the local density over right side of the string 
    with $\sigma \in (\pi/2,\pi)$ \cite{Horowitz:dt}.

    As was shown in \cite{Kluson:2002kk}, the
    solution (\ref{sol}) leads to a shifted form
    of the BRST operator
    \begin{equation}\label{newBRSTK}
    \tilde{Q}(X)=
    Q(X)+\Phi_0\star X-(-1)^{|X|} X\star \Phi_0=
    e^{-K}\left( Q(e^{K}(X))\right) \ 
    \end{equation}
    in the action for the fluctuation field when
    we perform an expansion of string field $\Phi$ 
    around the classical solution  $\Phi_0$ in
    (\ref{SFTaction}). In (\ref{newBRSTK}) 
    the symbol  $|X|$  denotes the ghost number of
    the string field $\Phi_0$.

    The previous results were used in
    \cite{Kluson:2002ex,Kluson:2002hr}.
     In the 
    next section we will consider another
     form
    of the operator $K$, one that resembles the 
    form of the generator of dilation in
    the conformal field theory. 

    \section{Scaling transformation}
    Let us start with the world-sheet action
    for the  bosonic string  
    \begin{equation}\label{stringaction}
    S=-\frac{1}{4\pi\alpha'}\int 
    d^2\sigma \eta^{\alpha\beta}\partial_{\alpha}
    X^{K}\partial_{\beta}X^{L}\eta_{LK} \ ,
    \end{equation}
    where $\eta^{\alpha\beta}$
    is two-dimensional Minkowski metric with
    signature $\eta_{\alpha\beta}=diag(-1,1)$  and $\eta_{KL}, 
    K, L=0,\dots,25$
    is a 26-dimensional target space Minkowski metric
    with signature $(-,+,\dots,+)$.

     From the action (\ref{stringaction}) we obtain the
     momentum conjugate to $X^{K}(\sigma)$
    \begin{equation}
    P_{K}(\sigma)=\frac{\delta L}
    {\delta \dot X^{K}(\sigma)}=
    \frac{1}{2\pi\alpha'}
    \dot X^{L}(\sigma)\eta_{LK} 
    \end{equation}
    with the following commutation relation
    \begin{equation}\label{com}
    [P_{K}(\sigma),X^{L}(\sigma')]=
    -i\delta (\sigma,\sigma') \delta_{K}^{L} \ ,
    \end{equation}
    where $\delta (\sigma,\sigma')$ is a delta function
    that  obeys Neumann 
    boundary conditions so that can be written as
    \begin{equation}
    \delta (\sigma,\sigma')=\frac{1}{\pi}
    \sum_{n=-\infty}^{\infty}\cos n\sigma
    \cos n \sigma'\ .
    \end{equation}
    We take the ghost
    number zero operator $K$ to be of the form
    \footnote{In our convention:
    $K,L=0,\dots,25 \ , i,j=0,\dots, P \ ,
    \mu,\nu=P+1,\dots,25 $ .}
    \begin{equation}
    K=i
    \int_0^{\pi}d\sigma X^i(\sigma)P_{i}(\sigma) \ ,
    \end{equation}
    where we do not care about a possible constant
    factor arising from the fact that ordering of
    the operators is not unique. E.g. starting from a  symmetric 
    ordering description and
    using (\ref{com}) would give the above $K$ multiplied by a constant which      we discard.
    For later use we calculate the following 
     commutators
    \begin{eqnarray}\label{comk}
    [P_KP_L(\sigma)\eta^{KL},\epsilon 
    K]=2\epsilon
    P_iP_j(\sigma)\eta^{ij} , \nonumber \\
    \left[X'^KX'^L(\sigma)\eta_{KL},\epsilon K\right]=
    -2\epsilon \eta_{ij}X'^iX'^j(\sigma) ,
    \nonumber \\
    \left[X'^KP_K(\sigma),\epsilon K\right]=
    0 \ . \nonumber \\
    \end{eqnarray}
    Start with $\epsilon <<1$ we
    have
    \begin{equation}\label{phi0}
    \Phi_0=e^{-K_L(\mathcal{I})}\star Q
    (e^{K_L(\mathcal{I})})\sim
    (\mathcal{I}-K_L(\mathcal{I}))\star Q
    (\mathcal{I}+K_L(\mathcal{I}))=
    Q(K_L(\mathcal{I})) 
    \end{equation}
    which can be written as
    \begin{eqnarray}
    \Phi_0=Q(K_L(\mathcal{I}))=Q_L(K_L(\mathcal{I}))+
    Q_R(K_L(\mathcal{I}))=\nonumber \\
    =Q_L(K_L(\mathcal{I}))+K_L(Q_R(\mathcal{I}))=
    Q_L(K_L(\mathcal{I}))-K_L(Q_L(\mathcal{I}))=
    [Q_L,K_L](\mathcal{I}) \  \nonumber \\
    \end{eqnarray}
    using \cite{Horowitz:dt}
    \begin{equation}
    Q_L(\mathcal{I})=-Q_R(\mathcal{I}) \ .
    \end{equation}
    Furthermore
    \begin{eqnarray}
    Q_R(K_L(\mathcal{I}))=\mathcal{I}\star
    Q_R(K_L(\mathcal{I}))=-Q_L(\mathcal{I})\star
    K_L(\mathcal{I})=\nonumber \\
    =Q_L(\mathcal{I})\star K_R(\mathcal{I})=
    -K_L(Q_L(\mathcal{I}))\star \mathcal{I}=
    -K_L(Q_L(\mathcal{I})) \ .
    \nonumber \\
    \end{eqnarray}
    Let us finally define $D_L(\mathcal{I})$ by
    \begin{equation}
    \Phi_0=D_L(\mathcal{I})=[Q_L,K_L](\mathcal{I}) \ .
    \end{equation}
    We then obtain the quadratic
    term for the fluctuation fields $\Psi$ in
    the form
    \begin{eqnarray}\label{Q'}
    \int \Psi\star Q'\Psi=\int \Psi
    \star Q(\Psi)+\int \Psi\star D_L(\mathcal{I})\star
    \Psi+
    \int \Psi\star\Psi \star D_L(\mathcal{I})= \nonumber \\
    =\int \Psi\star Q(\Psi)-\int \Psi\star D(\Psi)
    \nonumber \\
    \end{eqnarray}
    where we used
    \begin{eqnarray}
    \int \Psi\star D_L(\mathcal{I})\star
    \Psi=-\int \Psi \star \mathcal{I}
    \star D_R(\Psi)=-\int \Psi\star D_R(\Psi) \ , \nonumber \\
    \int \Psi\star \Psi\star D_L(\mathcal{I})=
    -\int \Psi\star \Psi\star D_R(\mathcal{I})=-
    \int \Psi\star D_L(\Psi) , \nonumber \\
    D_L(X)\star Y=-(-1)^XX\star D_R(Y) \ .\nonumber \\
    \end{eqnarray}
    The last formula may be proven as in
    \cite{Horowitz:dt} using the fact that the  commutator
    $[Q,K]$ has odd grading.
    Let us calculate this commutator. Since we have
    \begin{eqnarray}
    Q=\frac{1}{\pi}\int_0^{\pi}d\sigma J_{0}(\sigma)=
    \frac{1}{\pi}\int_0^{\pi}
    d\sigma c^{\alpha}(\sigma)T_{\alpha 0}(\sigma)+
    Q_{ghost} \ ,
    \nonumber \\
    T_{00}=T_{11}=\frac{1}{2}
    \left(4\pi^2\alpha' P_KP_L\eta^{KL}+
    \frac{1}{\alpha'}X'^KX'^L\eta_{KL}\right) \ ,
    \nonumber \\
    T_{10}=T_{01}=2\pi P_KX'^K \ , 
    X'^K=\partial_{\sigma}X^K \ , \nonumber \\
    \end{eqnarray}
    where $Q_{ghost}$ is the ghost contribution to 
    the BRST charge whose explicit form we will not
    require. 
    Using this form of the BRST operator we
    get
    \begin{equation}\label{Dc}
    D=[Q,K]=\frac{1}{\pi}\int_0^{\pi}d\sigma
    c^0(\sigma)\frac{1}{2}\left[
    4\pi^2\alpha'2\epsilon
    P_iP_j\eta^{ij}
    -2\epsilon 
    \frac{1}{
    \alpha'}
    X'^iX'^j\eta_{ij}\right]
     \ .
    \end{equation}
    Consequently the new BRST operator has the form
    \begin{eqnarray}\label{newQ1}
    Q'=Q-D=\frac{1}{\pi}
    \int_0^{\pi}d\sigma\left\{ c^0(\sigma)\
    \frac{1}{2}\left[
     4\pi^2\alpha'
    P_{\mu}P_{\nu}\eta^{\mu\nu}+
    \frac{1}{\alpha'}X'^{\mu}X'^{\nu}\eta_{\mu\nu}+
    \right.\right. \nonumber \\
    \left.\left.+
    4\pi^2\alpha'(1-2\epsilon)
    P_iP_j\eta^{ij}+
    (1+2\epsilon)\frac{1}{\alpha'}
    X'^iX'^j\eta_{ij})
    \right]+\right. \nonumber \\
    +\left. 
    c^1(\sigma) 2\pi X'^KP_K(\sigma)\right\}+
    Q_{qhost} \ . \nonumber \\
    \end{eqnarray}
    We see   from 
    (\ref{newQ1}) that  the 
    numerical factors in front of 
    $P_{i}P_j , X'^iX'^j $ now depend on
    the  parameter $\epsilon$. There is a
    suggestive physical interpretation of
    this phenomena which transpires after an analysis of the case
    of finite $\epsilon$. To this end, we calculate exact form of the new
    BRST operator given in (\ref{newBRSTK}).
    To do this calculation we define the 
    function
    \begin{equation}
    F(t)=e^{-\epsilon K t}\left(Q(e^{\epsilon Kt})
    \right) , \ F(t=0)=Q \ , F(1)=\tilde{Q} \ .
    \end{equation}
    It is easy to see that 
    \begin{equation}\label{Qtilde}
    \tilde{Q}=
    F(0)+\sum_{n=1}^{\infty}
    \left.\frac{1}{n!}\frac{d^nF}{d^nt}t^n\right|_{t=1}=
    Q+\sum_{n=1}^{\infty}
    \frac{(-1)^n \epsilon^n}{n!}
    [K,[K,\dots,[K,Q]]] \ .
    \end{equation}
    so to determine 
    (\ref{Qtilde}) we must calculate all
    commutators in  (\ref{Qtilde}).
    The second order commutator is equal to
    \begin{eqnarray}
    \left[K,[K,Q]\right]=-[K,D]=[D,K]=
    \nonumber \\
    =\frac{1}{2\pi}\int_0^{\pi}
    d\sigma c^0(\sigma)\frac{1}{2}
    \left(4\pi^2\alpha' 2^2
    P_iP_j\eta^{ij}+
    2^2\frac{1}{\alpha'}
    X'^iX'^j\eta_{ij}
    \right) \ , \nonumber \\
    \end{eqnarray}
    where we have used 
    (\ref{Dc}) and (\ref{comk}). We
    conclude that the new BRST operator will have
   the form
    \begin{eqnarray}\label{Qnew}
    Q'=
    \frac{1}{\pi}
    \int_0^{\pi}d\sigma\left\{ c^0(\sigma)\
    \frac{1}{2}\left[\left(
    4\pi^2\alpha' P_{\mu}P_{\nu}\eta^{\mu\nu}
    +\frac{1}{\alpha'}X'^{\mu}X'^{\nu}
    \eta_{\mu\nu}\right)
     +\right.\right. \nonumber \\
    \left.\left.+    
    (1-2\epsilon+
    \frac{1}{2}(2\epsilon)^2+\dots
    )
    4\pi^2\alpha'P_iP_j\eta^{ij}+\right.\right.
    \nonumber \\
    \left.\left.+
    (1+2\epsilon+\frac{(2\epsilon)^2}{2}
    +\dots)\frac{1}{\alpha'}
    X'^iX'^j\eta_{ij}\right]
    +c^1(\sigma)2\pi X'^KP_K(\sigma)\right\}+
    Q_{qhost}= \nonumber \\
    =\frac{1}{\pi}
    \int_0^{\pi}d\sigma\left\{ c^0(\sigma)\
    \frac{1}{2}\left[
    4\pi^2\alpha' P_{\mu}P_{\nu}\eta^{\mu\nu}
    +\frac{1}{
    \alpha'}X'^{\mu}X'^{\nu}\eta_{\mu\nu}+\right.\right.
     \nonumber \\
    \left.\left.
    +
    e^{-2\epsilon}4\pi^2\alpha'P_iP_j\eta^{ij}+
    e^{2\epsilon}\frac{1}{\alpha'}
    X'^iX'^j\eta_{ij}\right]
    +c^1(\sigma)2\pi  X'^KP_K(\sigma)\right\}+
    Q_{qhost} \ .\label{320}
    \nonumber \\
    \end{eqnarray}
    In the next section we try to interpret
    this solution. 
    \section{Interpretation of the solution}
    In this section we interpret the solution (\ref{320}). 
    As we have seen, the string field theory action
    expanded around the new solution has the form
    \begin{equation}
    S=-\frac{1}{2}\int \Psi\star Q'
    \Psi -\frac{1}{3}\int \Psi\star\Psi\star \Psi
    \end{equation}
    with 
    \begin{eqnarray}\label{Qnew2}
    Q'=
    \frac{1}{\pi}
    \int_0^{\pi}d\sigma\left\{ c^0(\sigma)\
    \frac{1}{2}\left[\left(
    4\pi^2\alpha'P_{\mu}P_{\nu}\eta^{\mu\nu}
    +\frac{1}{
    \alpha'}X'^{\mu}X'^{\nu}\eta_{\mu\nu}\right)+
    \right.\right. \nonumber \\
    \left.\left.+
    \left(e^{-2\epsilon}4\pi^2\alpha'P_iP_j\eta^{ij}+
    e^{2\epsilon}\frac{1}{\alpha'}
    X'^iX'^j\eta_{ij}\right)\right]+\right.\nonumber \\
    \left.+c^1(\sigma)2\pi  X'^KP_K(\sigma)\right\}+
    Q_{qhost} \ .
    \nonumber \\
    \end{eqnarray}
        In what follows we will discuss mainly the shifted
    part of the BRST operator. We also introduce
    the parameter
    \begin{equation}
    e^{\epsilon}=t
    \end{equation}
    We see that for $t=1$ we regain the original action.
    For $t^2\rightarrow 0$ we rewrite the BRST operator
    as 
    \begin{eqnarray}\label{Qshift}
    Q'=\frac{1}{\pi}
    \int_0^{\pi}d\sigma\left\{ c^0(\sigma)
    \frac{1}{2}\left[
    e^{-2\epsilon}4\pi^2\alpha'P_iP_j\eta^{ij}+
    e^{2\epsilon}\frac{1}{\alpha'}
    X'^iX'^j\eta_{ij})\right]
    +c^1(\sigma)2\pi X'^iP_i(\sigma)\right\}=
    \nonumber \\
    =
    \frac{1}{\pi}
    \int_0^{\pi}d\sigma\left\{ c^0(\sigma)
    \frac{1}{2}\left[
    4\pi^2\frac{\alpha'}{t^2}P_iP_j\eta^{ij}+
    \frac{t^2}{\alpha'}
    X'^iX'^j\eta_{ij})\right]
    +c^1(\sigma)2\pi X'^iP_i(\sigma)\right\}=
    \nonumber \\=
    \frac{1}{\pi}
    \int_0^{\pi}d\sigma\left\{ c^0(\sigma)
    \frac{1}{2}\left[
    4\pi^2\alpha'_{eff}P_iP_j\eta^{ij}+
    \frac{1}{\alpha'_{eff}}
    X'^iX'^j\eta_{ij})\right]
    +c^1(\sigma)2\pi X'^iP_i(\sigma)\right\}
    \ , \nonumber \\  
    \end{eqnarray}
    where we have introduced 
    \begin{equation}
    \alpha'_{eff}=\frac{\alpha'}{t^2} \ .
    \end{equation}
    Now we have
    \begin{eqnarray}
    t^2\rightarrow \infty \Rightarrow
    \alpha'_{eff}\rightarrow 0 \ , \nonumber \\
    t^2\rightarrow 0 \Rightarrow
    \alpha'_{eff}\rightarrow \infty \ .  \nonumber \\
    \end{eqnarray}
    The first limit corresponds to the
    zero slope limit in which case the ordinary string
    action reduces to the action for massless particle.
    It is remarkable that in our solution 
    this limit can be anisotropic.
    The second limit corresponds to the tensionless limit
    \cite{Karlhede:wb,
     Lindstrom:1990qb,Isberg:1992ia,Isberg:1993av,
    Sundborg:2000wp,Haggi-Mani:2000ru}.

    To see this more directly, let us follow
    \cite{Tseytlin:2002gz,Nastase:2000za,Lindstrom:2000pp,
    Svendsen:1999tp}.
    The first  limit $\alpha_{eff}'\rightarrow 0$ is the well
    known zero slope limit where the string effectively collapses
    to a point. From (\ref{Qshift}) it is clear
    that to have a finite expression the terms proportional
    to $X'$ should be equal to zero, in other words we consider
    the zero modes of the field on the world-sheet only.
    As a result (\ref{Qshift}) reduces to the BRST operator for
    massless particle.

    In the second limit $\alpha'_{eff}\rightarrow \infty$ 
    the term proportional to $\frac{1}{\alpha'_{eff}}
    X'^iX'^j$ goes to zero. We must also rescale
    $2\pi\sqrt{\alpha'_{eff}}P_i=
    \mathcal{P}_i$ to ensure that the energy density is
    finite. As a result,we obtain the following shifted BRST operator
    \begin{equation}\label{BRSTtensionless}
    Q_{shifted}=\frac{1}{\pi}
    \int_0^{\pi}d\sigma\left\{ c^0(\sigma)
    \frac{1}{2}
    \mathcal{P}_i\mathcal{P}_j\eta^{ij}
    \right\}
      \ .
    \end{equation}
    which is the BRST operator for an infinite collection
    of independent massless pointparticles   
    as we expect in for
    tensionless strings 
     \cite{Karlhede:wb,Lindstrom:1990qb,Isberg:1992ia,Isberg:1993av,
    Sundborg:2000wp,Haggi-Mani:2000ru}.
    We must also mention that the th tensionless
    limit $\alpha'\rightarrow \infty$ was also studied in
    very interesting paper \cite{Chu:2002mg} with the
    similar results.

    \section{Conclusion and open questions}\label{fifth}
    As we have seen we can easily find solutions of
    the string field theory equation of motion that
    depends on one single parameter
    that can be interpreted as the string effective length scale.
    In our solution this length scale can  vary continously
    from $0$ that corresponds to the zero slope limit
    in ordinary string theory to $\alpha'_{eff}=\infty$ 
    corresponding to the tensionless limit of the string
    theory. Generally, this effective string tension
    depends on the orientation of the solution, which is a similar
    effect to that for the effective tension in NCOS string as discussed
    in \cite{Seiberg:2000ms,Gopakumar:2000na,
    Gomis:2000zz,Aharony:2000gz,Gopakumar:2000ep,
    Bergshoeff:2000ai,Gomis:2000bd,
    Danielsson:2000gi,Klebanov:2000pp,
    Lu:2001qf,Danielsson:2000mu,Kristiansson:2000xv,
    Garcia:2002fa}.
    We hope that our result could be helphul for 
    the study the relation between tensile and tensionless
    strings
    \cite{Karlhede:wb,Lindstrom:1990qb,Isberg:1992ia,Isberg:1993av,
    Sundborg:2000wp,Haggi-Mani:2000ru}.
    It would be very interesting to study the conformal
    field theory on the string world-sheet action with
    an-isotropic tension. We hope to return to this problem
    in future.

    {\bf Acknowledgement}
    It is  my pleasure to thank all people at Departmant
    of Theoretical Physics in Uppsala, especially to
    Ulf Lindstrom for very stimulating discussions and
    to John Minahan, Ulf Danielsson 
    support in my work. This work is partly supported 
    by EU contract HPRN-CT-2000-00122. This work is also
    supported by the Czech Ministry of Education under
    Contract No. 143100006.

    
    \end{document}